\documentclass[final,5p,times,twocolumn]{elsarticle}
\usepackage{graphicx}
\usepackage{amssymb}

\begin{document}

\begin{frontmatter}

\title{A dedicated algorithm for calculating ground states for the triangular random bond Ising model}

\author{O. Melchert}
\ead{oliver.melchert@uni-oldenburg.de}

\author{A. K. Hartmann}
\ead{alexander.hartmann@uni-oldenburg.de}

\address{
Institut f\"ur Physik, Universit\"at Oldenburg, 
Carl-von-Ossietzky Stra\ss{}e 9--11, 26111 Oldenburg, Germany} 

\begin{abstract}
In the presented article we present an algorithm for the computation of
ground state spin configurations for the $2d$ random bond Ising model on planar 
triangular lattice graphs.
Therefore, it is explained how the respective ground state problem can be mapped to an auxiliary 
minimum-weight perfect matching problem, solvable in polynomial time. 
Consequently, the ground state properties as well as
minimum-energy domain wall (MEDW) excitations for very large $2d$ systems, e.g.\ lattice 
graphs with  up to $N\!=\!384\!\times\!384$ spins, can be analyzed very fast.

Here, we investigate the critical behavior of the
corresponding $T\!=\!0$ ferromagnet to spin-glass transition, signaled 
by a breakdown of the magnetization, using finite-size scaling analyses of 
the magnetization and MEDW excitation energy and we contrast our numerical 
results with previous simulations and presumably exact results.
\end{abstract}


\end{frontmatter}

\section{Introduction}
\label{sect:Intro}
Triggered by the exchange of ideas between 
computer science and theoretical physics in the past decades,
it was realized that several basic problems in the context of 
disordered systems relate to ``easy'' optimization
problems. These are problems where the solution time is polynomial in the size of 
the problem description.
As a result, many disordered systems can now be 
analyzed numerically exact through computer simulations
by using fast combinatorial optimization algorithms \cite{bastea1999,opt-phys2001,rieger2002}.
E.g., ground state (GS) spin configurations for the random-field Ising magnet
(in any dimension $d$) can be obtained by computing the maximum flow for an 
auxiliary network problem \cite{opt-phys2001}. 
Another example is the $2d$ Ising spin glass (ISG), where the lattice can be 
embedded in a plane. For this model, the problem of finding a GS 
spin configuration for a given realization of the nearest neighbor couplings 
can be mapped to an appropriate minimum-weight perfect-matching (MWPM) problem 
\cite{opt-phys2001,SG2dReview2007}.
Finally, the MWPM
problem can be solved in polynomial time by means of exact combinatorial 
optimization algorithms \cite{cook1999}.
Thus, the planar $2d$ ISG can be studied directly at 
zero temperature without equilibration problems and within polynomial time.
Hence, very large systems can be considered, giving very precise and reliable
estimates for the observables.
Actually, there are different approaches that allow for an exact computation of 
GSs for the planar $2d$ ISG \cite{bieche1980,barahona1982,pardella2008}.
Albeit all of these approaches rely on the computation of MWPMs on an auxiliary graph, 
they differ regarding the subtleties of the mapping to the respective auxiliary problem.
The most efficient of these approaches (see Ref.\ \cite{pardella2008}) is based on 
the Kasteleyn treatment of the Ising model \cite{kasteleyn1963}, which previously was 
also used to obtain extended ground states for the $2d$ ISG with fully periodic 
boundary conditions \cite{thomas2007}.

\begin{figure}[t!]
\centerline{
\includegraphics[width=1.0\linewidth]{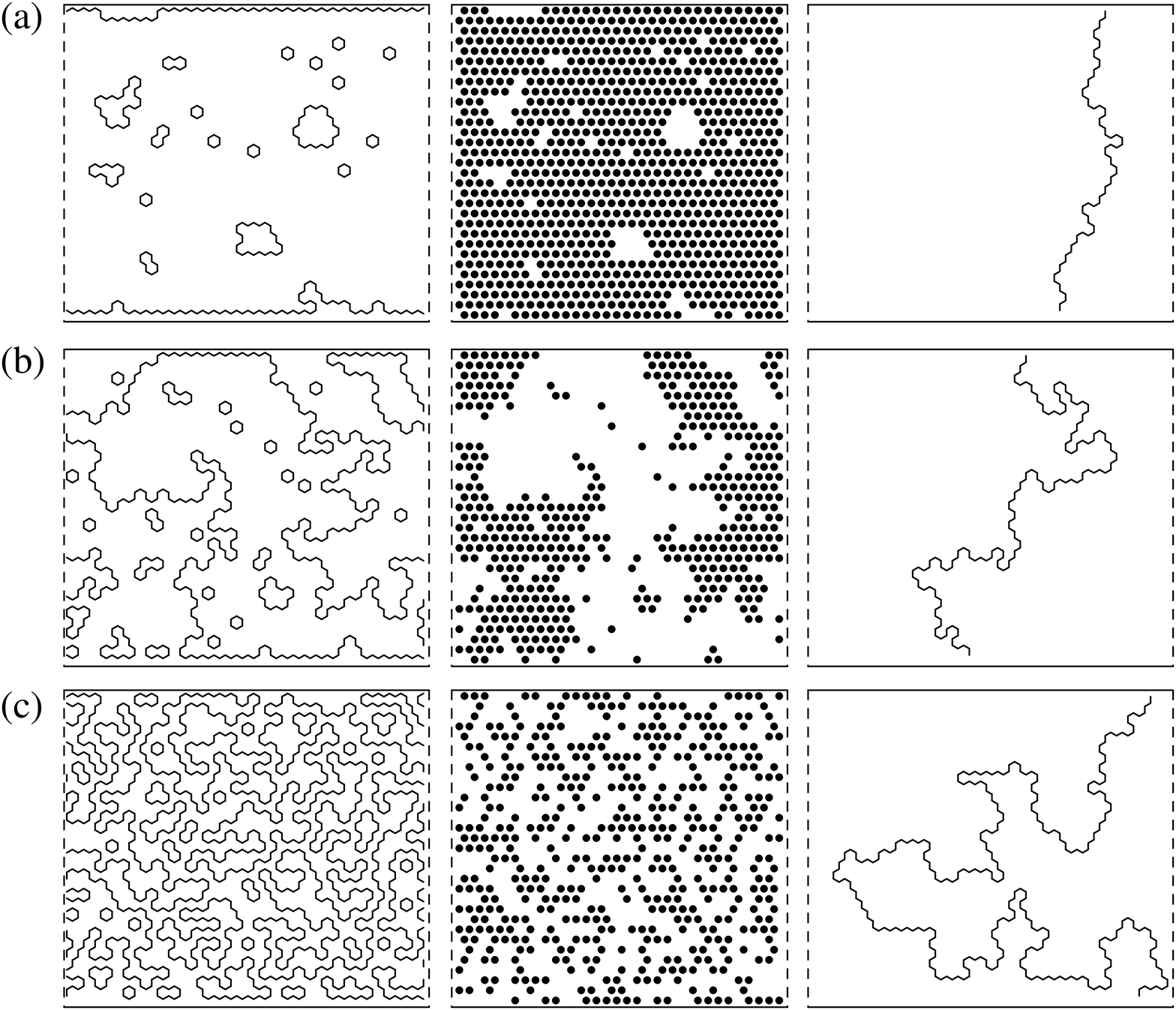}} 
\caption{
Samples of a $\pm J$ random bond Ising spin system on 
a triangular lattice of side length $L\!=\!32$. 
The samples are taken at three different values of the disorder
parameter $p$.
(a) $p\!=\!0.15$ characterized by a ferromagnetic GS,
(b) $p\!=\!0.2$ characterized by a GS with SG order and
(c) $p\!=\!0.5$, i.e. the canonical $\pm J$ ISG.
In the figure, 
periodic BCs are indicated by the dashed vertical lines. 
From left to right: 
(Left) Transition graph that describes the difference between a 
ferromagnetic reference spin configuration and the GS spin 
configuration, 
(center) corresponding GS, 
(right) MEDW excitation relative to the GS.
\label{fig1}}
\end{figure}

Here, we introduce a dedicated algorithm that yields exact GS spin configurations for 
the $2d$ random-bond Ising model (RBIM) on planar triangular lattice graphs.
As the previous approaches, the algorithm presented here requires to solve an 
associated MWPM problem. The corresponding mapping uses a relation between perfect 
matchings and paths on a graph \cite{ahuja1993,melchertThesis2009}. In effect, these paths can be used to 
partition the graph into domains of up and down spins that comprise a GS spin configuration, 
see Fig.\ \ref{fig1}. Consequently, the GS properties as well as minimum-energy domain wall 
(MEDW) excitations, see Fig.\ \ref{fig1}, can be analyzed very fast.
The presented algorithm enables us to study large systems, while allowing for an appropriate 
disorder average within a reasonable amount of computing time.
In this regard, it requires to compute a MWPM for an auxiliary graph with $O(N)$ edges only (wherein 
$N$ is the number of spins on the lattice).
However, note that the algorithm presented
here is asymptotically not faster than the algorithm presented in Refs.\ \cite{pardella2008,thomas2007},
but it highlights the algorithmic relation between the GS problem for spin glasses and the recently
proposed negative-weight percolation (NWP) problem \cite{melchert2008}.

In the presented article, we investigate the critical behavior of the $T\!=\!0$ ferromagnet (FM) to spin-glass (SG) 
transition for the $2d$ RBIM, signaled by a breakdown of the magnetization, using 
finite-size scaling (FSS) analyses of the MEDW excitation energy.
In this regard, we obtain a highly precise estimate of the critical point
for the triangular lattice geometry and we verify the critical exponents
obtained earlier for the RBIM on the planar square lattice \cite{amoruso2004,melchert2009}.
Finally, we contrast our numerical results with previous simulations 
and presumably exact results \cite{bendisch1997}.

The remainder of the presented article is organized as follows.
In section \ref{sect:model}, we introduce the model in more detail and 
we outline the algorithm used to compute the GS spin configurations.
In section \ref{sect:results}, we present the results of 
our numerical simulations and in section \ref{sect:conclusions} we 
conclude with a summary.

\section{Model and Algorithm}
\label{sect:model}
%
\begin{figure*}[t!]
\centerline{
\includegraphics[width=0.8\linewidth]{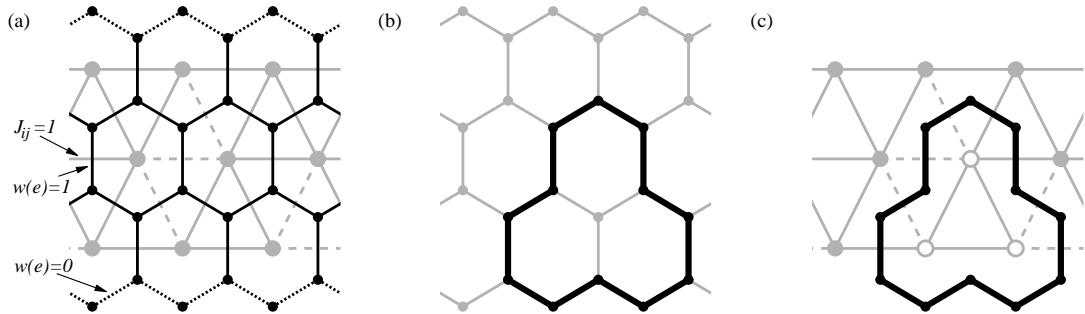}} 
\caption{
Illustration of the computation of the transition graph that 
allows to determine a GS spin configuration on a planar triangular RBIM.
The figure illustrates a sample system of side length $L=3$ and periodic
BCs in the horizontal direction.
(a) Mapping of the original lattice $G$ (grey edges, triangular geometry),
    where a solid (dashed) line indicates a ferromagnetic (antiferromagnetic)
    bond, to the weighted dual graph $G_D$ (black edges, honeycomb geometry). 
    Note that additional edges (dotted lines) where introduced to
    account for the open boundary conditions in the respective direction. These
    edges carry zero weight.
(b) Minimum weight set of loops (bold black edges) on $G_D$ as obtained from a mapping
    to the NWP problem (not shown here, see Ref.\ \cite{melchert2008}).
(c) Loop on the dual surrounding a cluster of spins on the spin lattice. 
    If the orientation of the spins is chosen as explained in the text, this
    procedure yields a GS spin configuration. 
    In the figure, spin orientations are distinguished by gray filled and non-filled
    circles.
\label{fig2abc}}
\end{figure*}  

In the presented article we perform GS calculations for the $2d$ RBIM,
where the respective model consists of $N=L \times L$ Ising spins
$\sigma=(\sigma_1,\ldots,\sigma_N)$, where $\sigma_i=\pm 1$, located on
the sites of a planar triangular lattice graph. Therein, the energy of a given spin
configuration is measured by the Edwards-Anderson Hamiltonian
\begin{eqnarray}
H(\sigma) = -\sum_{\langle i,j \rangle} J_{ij} ~\sigma_i \sigma_j,
\label{eq:EA_hamiltonian}
\end{eqnarray}
where the sum runs over all pairs of nearest-neighbor spins (on the triangular
lattice) with periodic boundary conditions (BCs) in the $x$-direction and free BCs 
in the $y$-direction.  
In the above energy function, the bonds $J_{ij}$ are quenched random variables 
drawn from the disorder distribution
\begin{eqnarray}
P(J)\!&=&\! p~\delta(J\!+\!1)~+~(1\!-\!p)~\delta(J\!-\!1).
\end{eqnarray}
Therein, one realization of the disorder consists of a random
fraction $p$ of antiferromagnetic bonds ($J=-1$) that prefer an antiparallel alignment of 
the coupled spins, and a fraction $(1-p)$ of ferromagnetic bonds ($J=1$) in 
favor of parallel aligned spins.
In general, the competitive nature of these interactions gives rise to frustration. 
A plaquette, i.e.\ an elementary triangle on the lattice, is said to be frustrated if it is
bordered by an odd number of antiferromagnetic bonds. 
In effect, frustration rules out a GS (i.e.\ a minimizer $\sigma_{GS}$ of Eq.\ \ref{eq:EA_hamiltonian}) in which all the bonds are satisfied. 
As limiting cases one can identify the Ising ferromagnet at
$p=0$ and the canonical $\pm J$ ISG at $p=1/2$. 
Hence, as a function of the disorder parameter $p$ we expect to find a ferromagnetic 
phase (spin-glass phase) for $p\!<\!p_c$ ($p\!>\!p_c$), wherein $p_c$ denotes
the critical point at which the $T\!=\!0$ FM-SG transition takes place.
For the ISG with a bimodal disorder distribution the 
GS is highly degenerate and the average number of such GSs increases exponentially with $N$ \cite{saul1994,landry2002}.
Apart from the GSs, we here also aim to characterize the energetic properties of
MEDW excitations.
A domain wall is an interface that spans the system in the direction with thee 
free BCs. Now, the MEDW is such an interface with an excitation energy  
$\delta E$ that is minimal among all possible domain walls.
Due to the extensive degeneracy of the GSs, the ``lattice-path'' associated with a MEDW is 
not unique. I.e., there are many DWs with minimal excitation energy. Albeit the
geometric properties of a MEDW are not unique, its excitation energy is unique.
MEDWs for three different values of the disorder parameter $p$ are illustrated in Fig.\ \ref{fig1}.

We now give a brief description of the algorithm that we use to compute the GSs.
Therefore, we first set a reference spin configuration $\sigma_R$. The most 
convenient choice is a maximally polarized, i.e.\ ferromagnetic, configuration $\sigma_R=(+1,\ldots,+1)$. 
Then, we construct a weighted dual of the spin lattice as shown in Fig.\ \ref{fig2abc}(a).
Since the spin lattice considered here has a triangular geometry, the corresponding
dual graph possesses a honeycomb geometry.
Note, that we introduced $4L$ extra nodes on top and at the bottom of the dual in 
order to account for the free BCs along that direction and to maintain the honeycomb 
structure of the respective graph. 
This means, a triangular spin lattice of size $L\times L$ is transformed to a 
honeycomb lattice with an over all number of $(2 L)\times(L+1)$ nodes.
Hence, the topological dual graph associated to 
the triangular spin lattice is modified to some extend. 
Further note that adjacent 
extra nodes are connected by edges $e$ that carry a weight $\omega(e)=0$.
All other edges $e$ on the dual graph 
get an edge-weight $\omega(e)\!\equiv\! J_{ij} \sigma_{R,i} \sigma_{R,j}$. 
Therein, $e$ is assumed to cross a bond $J_{ij}$ on the spin lattice, where
$J_{ij}$ couples the two spins $\sigma_{R,i/j}$, see Fig.\ \ref{fig2abc}(a).
Consequently, the edge weight on the weighted dual is positive (negative), if the corresponding 
bond on the spin lattice is satisfied (broken) with respect to $\sigma_R$.
A pivotal observation is, that there exists an equivalence between clusters of adjacent spins on the spin
lattice that might be flipped in order to decrease the configurational energy of $\sigma_R$ and 
negative-weighted loops (i.e.\ closed paths) on the weighted dual graph.
In this regard, if a loop with negative weight on the dual is found, the cluster of spins surrounded
by this loop can be flipped so as to decrease the configurational energy of $\sigma_R$.
Finally, to obtain a GS spin configuration one needs to find a minimum-weight set of 
negative-weighted loops on the dual graph (see discussion below). This set of loops comprises the transition graph for the 
given realization of the disorder, as illustrated in Fig.\ \ref{fig2abc}(b).
Since initially a ferromagnetic reference configuration was chosen, 
the GS is obtained if the orientation of the spins on the spin lattice is chosen 
such, that 
(i)  spins within a cluster are aligned in the same direction and
(ii) spins in adjacent clusters are aligned in opposite directions. 
The resulting GS is indicated in Fig.\ \ref{fig2abc}(c), see also Fig.\ \ref{fig1}.

Here, for the $2d$ RBIM on a planar triangular lattice, where the dual has a honeycomb
geometry, the minimum weight set of loops on the weighted dual can be obtained 
by means of a mapping to the NWP problem, as explained in \cite{melchert2008}.
In brief, the NWP statement consists in the task to find
a minimum-weight set of nonintersecting negative-weighted loops for a given weighted graph. 
Therefore, it considers a minimum-weight perfect matching problem on an 
associated auxiliary graph with $O(N)$ edges (provided that the input graph
has $O(N)$ edges, as it is the case here), from which the set of loops can be deduced.
Note that the mapping to the NWP problem yields the correct transition graph only
for this particular lattice setup, since any two-coloring of the spin lattice (i.e.\ assignment of up/down spin orientations)
can be composed by loops on the dual that do not intersect. That means, each site 
on the dual is an end-node of either 0 or 2 loop segments, as e.g.\ in Fig.\ \ref{fig2abc}(b).
In contrast, two-colorings of the spins on a square lattice might involve
loops on the dual that involve figure-8 twists. That means, each site on
the dual is end-node of either 0, 2 or 4 loop segments. 
For the latter problem, a different mapping \cite{pardella2008} was 
used recently to obtain exact GSs for $2d$ ISGs on a 
square lattice with free BCs in at least one direction within 
polynomial time. This mapping was further used to compute 
``extended'' GSs for the $2d$ ISG with fully periodic BCs \cite{thomas2007}. 

Now, the interpretation of the $T\!=\!0$ FM-SG transition in terms of the 
NWP problem reads as follows: For small values of $p$, there are only 
few bonds on the spin lattice that are not satisfied by the reference 
spin configuration. Accordingly, there are only few small loops that 
comprise the transition graph. For all nonzero values of $p$, 
a sufficiently large lattice will feature at least some small loops that
surround an elementary plaquette on the dual. These small loops correspond
to local ``manipulations'' of the order parameter (i.e.\ the magnetization), only.
Hence, in the thermodynamic limit, the GS has still ferromagnetic order (see Fig.\ \ref{fig1}(a)).
However, if the value of $p$ increases and exceeds a critical value $p_c$,
large loops appear that have a linear extension of the order of the system size
and eventually span the system along the direction with the periodic boundary conditions.
These loops represent global manipulations of the order parameter, that, in the
thermodynamic limit, destroy the ferromagnetic order of the GS (see Figs.\ \ref{fig1}(b),(c)).

Once we obtained a GS spin configuration in this manner, we compute a MEDW by means of a similar
mapping, thoroughly explained in Ref.\ \cite{melchert2007}. 
In the following we will use the procedure outlined
above to obtain GSs and to investigate MEDWs for the RBIM 
introduced above.

\section{Results}
\label{sect:results}

\begin{figure}[t!]
\centerline{
\includegraphics[width=1.0\linewidth]{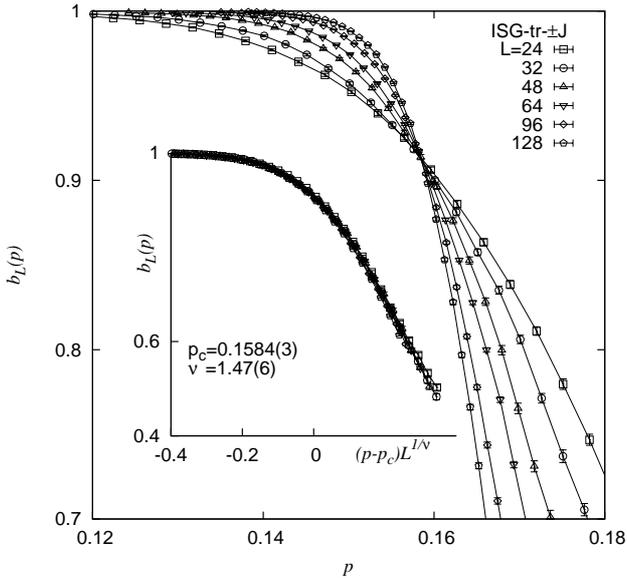}}
\caption{ \label{fig3}
Results of the finite-size scaling analysis for the binder parameter 
$b_L(p)$, considering different system sizes $L$.
The main plot shows the unscaled data close to the critical point and the insets illustrate the data 
collapse obtained after rescaling the raw data using the scaling
assumption discussed in the text and scaling parameters as listed in Tab.\ \ref{tab1}.}
\end{figure}

\begin{table}[b]
\caption{
Critical exponents for the $2d$ RBIM. From left to right: Problem setup
(SQ=square lattice, TR=triangular lattice), critical exponent of the 
correlation length $\nu$, order parameter exponent $\beta$, and
exponents $\phi_1$ and $\psi_1$ that characterize the scaling of the MEDW
excitation energy. The figures for SQ-a are taken from Ref.\ \cite{melchert2009}.
The figures for SQ-b are taken from Ref.\ \cite{amoruso2004}. \label{tab1}}
\begin{center}
\begin{tabular}[c]{lllll}
\hline
\hline
 Setup   & $\nu$     & $\beta$ & $\phi_1$ & $\psi_1$ \\
\hline
SQ-a     & $1.49(7)$ & $0.097(6)$ & $0.67(3)$ & $0.17(2)$ \\
SQ-b     & $1.55(1)$ & $0.09(1)$  & $0.75(5)$ & $0.12(5)$ \\
TR       & $1.47(6)$ & $0.086(5)$ & $0.68(8)$ & $0.15(2)$ \\
\hline
\hline
\end{tabular}
\end{center}
\end{table}
As pointed out above, at small values of $p$ there exists
an ordered ferromagnetic phase, while for large values of 
$p$ a spin-glass ordered phase appears. 
A proper order parameter to characterize the respective 
FM-SG transition is the magnetization per spin $m_L=|\sum_i \sigma_i|/ L^{2}$ 
for a system of linear extend $L$. 
Below, we perform a finite-size scaling analysis (FSS) in order
to locate the critical point $p_c$ and to estimate the
critical exponents that describe the scaling behavior of the magnetization 
in the vicinity of the critical point.
Therefore, we first consider the Binder parameter \cite{binder1981} 
\begin{eqnarray}
b_L\!=\!\frac{1}{2}\Big(3-\frac{\langle m_L^4\rangle}{\langle m_L^2\rangle^2 }\Big)
\end{eqnarray}
associated with the magnetization. It is expected to scale as
$b_L(p)\!\sim\!f_1[ (p-p_c) L^{1/\nu}]$,
wherein $f_1[\cdot]$ signifies a size-independent scaling function and
 $\nu$ denotes the critical exponent that describes the divergence of the correlation length as the 
critical point is approached. Here, we simulated triangular systems of side length $L=24$ through $128$
at various values of the disorder parameter $p$. Observables are averaged over 
$64\,000$ samples for the largest systems and we
used the data collapse generated by the scaling assumption above to obtain 
$p_c\!=\!0.1584(3)$ and $\nu\!=\!1.47(6)$ with a quality 
$S\!=\!0.94$ of the data collapse \cite{houdayer2004,autoScale2009}, see Fig.\ \ref{fig3}.
In general, the above scaling relation holds best near the critical point and 
one can expect that there are corrections to scaling off criticality. 
As a remedy, we restricted the latter scaling analysis to the interval 
$[-0.3,+0.3]$, enclosing the critical point on the rescaled abscissa. 

Further, the order parameter of the transition is expected to scale according to
the scaling relation
$\langle m_L(p)\rangle\!\sim\!L^{-\beta/\nu} f_2[(p-p_c) L^{1/\nu}]$,
where $f_2[\cdot]$ denotes a size-independent function,
and where the order parameter exponent $\beta$ can be obtained 
after fixing $\nu$ and $p_c$ to the values obtained from the analysis of the Binder 
parameter. 
The best data collapse ($S\!=\!1.01$) was obtained for the choice
$\beta\!=\!0.086(5)$ (not shown). 
\begin{figure}[t!]
\centerline{
\includegraphics[width=1.0\linewidth]{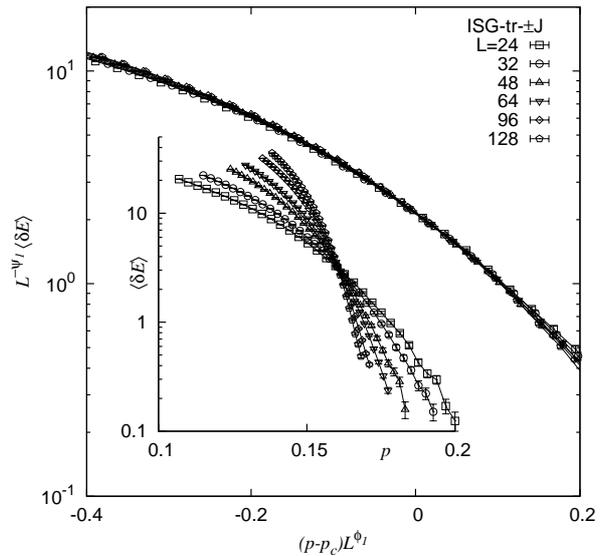}}
\caption{ \label{fig4}
Results of the finite-size scaling analysis for the average MEDW energy
$\langle \delta E \rangle$. 
The main plot shows the data collapse after rescaling the raw data using 
the scaling parameters listed in Tab.\ \ref{tab1} and the
inset illustrates the unscaled data close to the critical point.
}
\end{figure}  

Moreover, an analysis of the average MEDW excitation energy $\langle \delta E \rangle$ according
to the scaling assumption 
$\langle \delta E \rangle\!\sim\!L^{\psi_1} f_3[(p-p_c) L^{\phi_1}]$, see Ref.\ \cite{kawashima1997}, 
yields the critical point $p_c=0.1586(2)$, in agreement with the above estimate obtained 
using the Binder parameter. The critical exponents $\psi_1$ and $\phi_1$ are listed in Tab.\ \ref{tab1}.
Therein, we restricted the scaling analysis to the interval $[-0.1,+0.1]$ and obtained 
a best data collapse with $S\!=\!1.63$.

\begin{figure}[t!]
\centerline{
\includegraphics[width=1.0\linewidth]{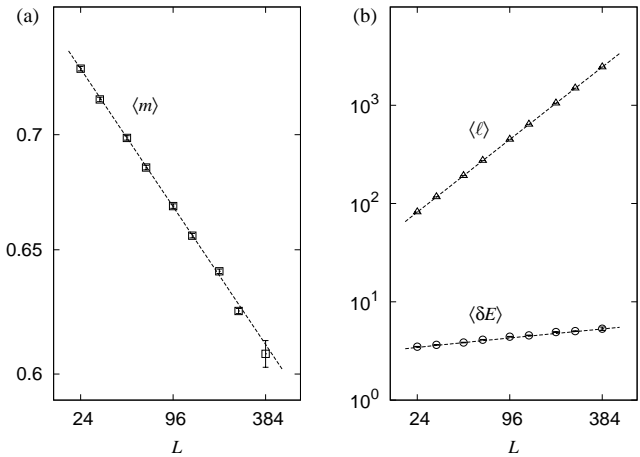}}
\caption{ \label{fig5ab}
Results of the finite-size scaling analysis at the critical point, 
where the $T\!=\!0$ FM-SG transition occurs.
(a) Scaling behavior of the magnetization $\langle m \rangle$, and, (b) scaling
of the average MEDW length $\langle \ell \rangle$ and the MEDW excitation energy $\langle \delta E \rangle$.
}
\end{figure}

Right at the critical point $p_c$ we performed additional simulations for spin lattices of up 
to $384\times 384$ spins (and $3.6\times 10^4$ samples), i.e.\ weighted dual graphs of up to $ 768 \times 385$ nodes. 
Upon analysis of the data we obtain the estimate $\beta=0.097(8)$ from the scaling behavior 
of the magnetization, see Fig.\ \ref{fig5ab}(a). We allowed for small deviations from a pure power-law scaling
using a scaling assumption of the form $\langle m \rangle\sim (L+\Delta L)^{-\beta/\nu}$, wherein $\Delta L = O(1)$.
Considering the scaling of the average MEDW excitation energy $\langle \delta E \rangle$ and using a similar scaling assumption as above, 
we found $\psi_1=0.15(1)$, see Fig.\ \ref{fig5ab}(b). 
Both these exponents agree within error bars with those obtained earlier, see Tab.\ \ref{tab1}.
As pointed out above, for the ISG with bimodal disorder, there a numerous MEDWs that differ 
regarding their geometric properties. However, here we also analyze the average length $\langle \ell \rangle$ 
of the particular MEDWs obtained within the simulations, see Fig.\ \ref{fig5ab}(b). Therefore, we considered a scaling 
according to the form $\langle \ell \rangle\sim (L+\Delta L)^{d_f}$, wherein $d_f$ signifies the 
fractal dimension of the MEDWs at $p_c$. We obtained $d_f=1.222(1)$ (and $\Delta L=O(1)$), which 
is in agreement with the value $d_f=1.222(1)$ found earlier for the $T\!=\!0$ FM-SG transition for 
the RBIM on a $2d$ square lattice, see Ref.\ \cite{melchert2009}.

\section{Conclusions}
\label{sect:conclusions}
In the presented article we have illustrated
how GSs for the $2d$ RBIM on planar triangular lattice graphs can be 
computed by a mapping to the NWP problem. 
I.e., the problem of finding a GS spin configuration for a
planar $2d$ triangular RBIM is equivalent to the NWP problem on a properly weighted
corresponding dual graph that exhibits a honeycomb structure.
Using this approach, we have investigated GSs and MEDW excitations
for the respective lattice structure.
Therein, 
a disorder parameter could be used to distinguish a ferromagnetic
and a spin-glass ordered phase. 
We characterized the corresponding $T\!=\!0$ FM-SG transition by means
of a FSS analysis of the magnetization and the MEDW excitation energy.

In this regard, we found that the values of the critical exponents obtained here agree within errorbars with those
obtained earlier for the $2d$ RBIM on a planar square lattice by 
considering a Gaussian bond distribution with ferromagnetic bias \cite{melchert2009} or a 
bimodal bond distribution \cite{amoruso2004}, as listed in Tab.\ \ref{tab1}.

Hence, the results for the triangular lattice structure obtained here highlights the
universality of the $T\!=\!0$ FM-SG transition.
Further, note that $p_c$ and $\nu$ found here agree well with the values $p_c\!=\!0.1583(6)$ and
$\nu\!=\!1.47(9)$ that characterize the negative-weight percolation of loops on $2d$ lattice 
graphs with a honeycomb geometry and fully periodic boundary conditions \cite{melchert2008}. 
Finally, the location of the critical point obtained here via FSS analysis is close to the
theoretical prediction $p_{c,tr}=0.15$, that was obtained for systems with fully periodic boundary 
conditions using the adjoined problem approach \cite{bendisch1994,bendisch1997}. 


\section*{Acknowledgment}
We are grateful to B. Ahrens for a critical reading of the manuscript.
OM acknowledges financial support from the VolkswagenStiftung (Germany)
within the program ``Nachwuchsgruppen an Universit\"aten''. 
The simulations were performed at the GOLEM I cluster for scientific 
computing at the University of Oldenburg (Germany).

\bibliographystyle{model1-num-names}
\bibliography{lit_RBIM.bib}
\end{document}